\documentclass[aps,amsmath,amssymb,reprint,superscriptaddress]{revtex4-1}

\usepackage{dcolumn}
\usepackage{bm}
\usepackage[utf8]{inputenc}
\usepackage[T1]{fontenc}
\usepackage{hyperref}
\usepackage[capitalize]{cleveref}
\usepackage{appendix}
\usepackage{cancel}
\usepackage{todonotes}
\usepackage{physics}
\usepackage{siunitx}
\usepackage[caption=false]{subfig}
\usepackage[bbgreekl]{mathbbol}
\usepackage{color,soul}

\DeclareSymbolFontAlphabet{\mathbbm}{bbold}
\DeclareSymbolFontAlphabet{\mathbb}{AMSb}
\graphicspath{{figures/}}

\newcommand{\eq}{{(0)}}
\newcommand{\noneq}{(\bm{v})}

\newcommand{\bV}{\bm{v}}
\newcommand{\bF}{\bm{F}}
\newcommand{\bX}{\bm{x}}

\newcommand{\REMOVE}[1]{}

\newcommand{\ahum}[1]{``#1''}
\newcommand{\olcite}[1]{Ref.~\cite{#1}}
\newcommand{\avg}[1]{\left< #1 \right>}

\newcommand{\ie}{{\it i.e.}}
\newcommand{\eg}{{\it e.g.}}

\begin{document}

\preprint{APS/123-QED}

%% OLD: Locally resolved dissipation in sliding friction on nanoscopic surfaces
\title{Spatially resolved atomic-scale friction: Theory and Simulation}

\author{Miru Lee}
\email{miru.lee@uni-goettingen.de}
\affiliation{Institute for Theoretical Physics, Georg-August-Universit\"at G\"ottingen, 37073 G\"ottingen, Germany}

\author{Richard L.C. Vink}
\email{rvink1@gwdg.de}
\affiliation{Institute of Materials Physics, Georg-August-Universit\"at G\"ottingen, 37073 G\"ottingen, Germany}

\author{Matthias Kr\"uger}
\email{matthias.kruger@uni-goettingen.de}
\affiliation{Institute for Theoretical Physics, Georg-August-Universit\"at G\"ottingen, 37073 G\"ottingen, Germany}

\date{\today}

\begin{abstract}We analyze the friction force exerted on a small probe particle sliding over an atomic-scale surface by means of a Green-Kubo relation and classical Molecular Dynamics simulations. We find that, on the atomic scale, the friction tensor can drastically vary as a function of position and sliding direction. The Green-Kubo relation yields this positional and directional dependence from equilibrium simulations of the {\it time dependent} covariance of force  acting on the probe. We find, unexpectedly, that the positional and directional dependence of energy dissipation is related to the (much simpler) {\it static} force covariance, especially in the limit where the probe only mildly perturbs the surface particles. In contrast, the (free) energy landscape experienced by the probe is in general {\it not} a good indicator of local dissipation. We also discuss optimization strategies making use of the locally and directionally resolved friction tensor. This enables us to find optimal sliding paths and velocity protocols, \eg, minimizing energy dissipation, between two points on the surface in a given time.\end{abstract}

\maketitle

\section{Introduction}

One of the \ahum{fruit fly} models of sliding friction is the Prandtl-Tomlinson 
(PT) model~\cite{10.1002/zamm.19280080202, 10.1080/14786440608564819}, which 
considers a single particle being dragged over a modulated potential landscape. 
Despite its simplicity, the PT-model captures certain aspects of atomic-scale 
friction quite well, the typical experimental realization being the tip of an 
atomic force microscope (AFM) moved over a surface~\cite{10.1038/374607a0, 
10.1098/rsta.2007.2164, 10.1021/cr960068q, persson2013sliding}. For the 
PT-model, the friction force required to maintain a constant sliding velocity 
is linearly proportional to the sliding velocity with the proportionality 
constant being the friction coefficient~\cite{10.1002/zamm.19280080202, 
muser11a,apostoli17a, 10.1103/physrevlett.84.1172, 
reimann04a, 10.1002/0471428019.ch5}. This linear relation holds provided that 
the sliding velocity is small, which is the so-called {\it linear response 
regime}. It has been found, in experiments, not only that the friction force 
depends on the direction of sliding~\cite{10.1103/physrevb.82.205417, 
10.1126/science.1113239}, but also that it varies in 
space~\cite{10.1103/physrevlett.97.136101, 10.1103/physrevlett.101.156102, 
10.1126/science.1150288, sasaki96a}. The observation that friction is spatially 
varying is quite remarkable because it implies that the friction coefficient is 
not simply a constant, but a function of position and sliding direction. For 
some sliding direction of interest, the friction coefficient appearing in the 
linear response relation should thus be regarded as a spatial average over all 
the positions that are sampled along the sliding path.

In this manuscript, we present theory and Molecular Dynamics simulations to 
unravel the full functional form of the friction coefficient (more accurately: 
a tensor) capturing the above mentioned spatial and directional dependencies. 
We do this for a small probe particle sliding over a crystalline model surface. 
Our theory is at the level of linear response, where, as it turns out, the 
positional and directional dependencies already markedly appear. The theory 
treats friction as being the manifestation of a {\it stochastic} process, as 
was already pointed out by~\citet{10.1002/zamm.19280080202}, and it naturally 
leads to a formulation in terms of stochastic Langevin 
equations~\cite{gauthier99a,muser11a,reimann04a,kantorovich08a,kantorovich08b}.

In line with this interpretation, we use the famous concept of Green-Kubo relations~\cite{kubo12a, zwanzig64a, kruger16a, maes06a, 10.1088/0034-4885/29/1/306,gauthier99a, dhont96a, zwanzig60a, zwanzig01a,bocquet97a,huang14a} to describe the linear response, \ie, the regime of small sliding velocities. Our approach is, in particular, parallel with the work of~\citet{gauthier99a}, employing a Green-Kubo relation to obtain energy dissipation originated from the motion of a probe.

The Green-Kubo relations allow to extract the local friction tensor from equilibrium thermal fluctuations of surface particles measured with the probe held at rest. This concept is found to be of practical advantage, producing results of higher statistical quality as compared to the \ahum{conventional} method of performing an explicit non-equilibrium simulation.

We find that, for the case of a probe particle sliding over an fcc $(111)$-surface, the friction tensor is a pronounced function of space and direction. This dependence can be understood from the mentioned Green-Kubo relations, which yield the friction tensor in terms of three characteristic properties: the frequency of fluctuations~\cite{PhysRevB.59.11777}, the relaxation time of fluctuations, and the covariance function of forces measured at equal times. Employing a weak coupling limit, where the probe is weakly coupled to the surface, we demonstrate that the former two hardly depend on the local position, so that the positional and directional variation of friction is entirely captured by the force-force covariance function at equal times. The latter is a {\it static} (as well as equilibrium) property of the system, and thus, independent of \eg, the mass of the surface particles. Intriguingly, we observe that the (free) energy landscape of the probe (a quantity that is experimentally accessible \cite{10.1103/physrevlett.97.136101}) is generally  {\it not} related to the local dependence of the friction tensor.

The Green-Kubo relations directly yield the {\it dissipative} parts of a sliding process, through which we address the question of the amount of dissipated energy under motion. The mentioned directional and positional dependencies thereby lend themselves to the possibility of optimizing the trajectory under the constraint of moving from one point to the other in a given amount of time. Such optimization concerns, on the one hand, the path, so that, under certain conditions, a detour is beneficial. On the other hand, it concerns the moving velocity, for which we derive conditions for minimal dissipation.

The paper is organized as follows. In~\cref{sec:theo}, we introduce the theoretical background and the rationale of our analysis. \Cref{sec:MD} explains our simulation setup and the relevant parameters. In~\cref{sec:results}, we first validate the locally resolved Green-Kubo relation, and then discuss the origin of the positional and directional dependence of friction. Thereafter, we present the optimization of energy dissipation by means of both path and velocity protocol. The concluding remarks can be found in~\cref{sec:conc}.

\section{Setup and theoretical background}\label{sec:theo}

\begin{figure}
\centering
\includegraphics[scale=1]{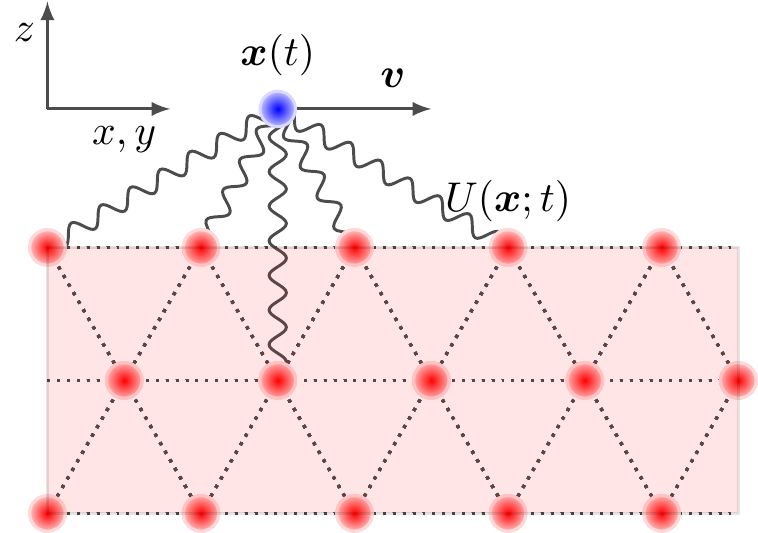}
\caption{A schematic of the system. The blue circle represents a probe particle at position $\bm{x}(t)$ that is moved with a prescribed velocity $\bm{v}:=\dv*{\bm{x}(t)}{t}$. The red circles represent substrate atoms which interact with each other via permanent nearest neighbor bonds. In addition, there is a short-ranged interaction between substrate atoms and probe particle. Since the substrate atoms thermally fluctuate about their equilibrium lattice positions, the interaction potential $U(\bm{x}(t);t)$ experienced by the probe is a stochastic variable.}
\label{fig:schematics}
\end{figure}

Our setup, schematically shown in \cref{fig:schematics}, consists of a probe being moved over a surface. The motion of the probe is prescribed, that is, the probe position $\bm{x}(t)$ as a function of time $t$ is controlled and given (the driving protocol). The probe interacts with the substrate atoms via a  potential, which is a function of the probe position and the positions of the substrate atoms. Due to thermal motion of the substrate atoms, the potential energy experienced by the probe, $U(\bm{x};t)$, explicitly depends on time in a stochastic manner. The key observable of interest is the force $\bm{F}$ felt by the probe due to the substrate atoms, 
\begin{equation}\label{eq:fdef}
  \bm{F}(t) = -\nabla_{\bm{x}}U(\bm{x};t)\left.\right|_{\bm{x}(t)}.
\end{equation}
In general, the resulting force depends on the details of the driving protocol. For simplicity, we consider, in what follows, a protocol whereby the probe moves with constant velocity parallel to the $xy$-plane (more general cases pose no additional principle challenges, as long as the driving velocity remains small). The system at time $t$ is then characterized by the position $\bm{x}(t)$ of the probe and its velocity $\bm{v}$. If the probe is at rest ($\bm{v}=\bm{0}$), the system will be in thermal equilibrium. We denote the average force by $\avg{\bm{F}}^{(0)}_{\bm{x}}$, which, in general, depends on the position $\bm{x}$. The connection to the corresponding free energy ${\cal F}(\bm{x})$ as a function of the probe position is then given by, 
\begin{align}\label{eq:eqf}
  \avg{\bm{F}}^{(0)}_{\bm{x}} = -\nabla_{\bm{x}}{\cal F}(\bm{x}).
\end{align}
For finite sliding velocity $\bm{v}$, the system is out of equilibrium, and the average force $\avg{\bm{F}}^{(\bm{v})}_{\bm{x}}$ will now differ from its equilibrium value. The force can be expanded in powers of the velocity, by means of the famous Green-Kubo relations~\cite{kubo12a, zwanzig64a, kruger16a, maes06a, 10.1088/0034-4885/29/1/306,gauthier99a, dhont96a, zwanzig60a, zwanzig01a}, which here read as,
\begin{equation}
\begin{split}
\left<\bm{F}\right>^{\noneq}_{\bm{x}} =& \left<\bm{F}\right>^\eq_{\bm{x}} \\
&- \beta \int_{0}^\infty \left<\bm{F}(t);\bm{F}(0)\right>^\eq_{\bm{x}} \dd{t} \cdot \bm{v}
+ \order{\bm{v}^2},
\end{split}
\label{eq:GK_F}
\end{equation}
with $\beta = 1/k_{\mathrm{B}}T$, temperature $T$, Boltzmann constant $k_\mathrm{B}$, and $\avg{\bm{A};\bm{B}} := \avg{\bm{A}\bm{B}} - \avg{\bm{A}}\avg{\bm{B}}$ the covariance tensor of the vectors $\bm{A}$ and $\bm{B}$. Note that \cref{eq:GK_F} relates the non-equilibrium force to the time dependent force covariance measured in equilibrium at the given position~$\bm{x}$.

The right hand side of \cref{eq:GK_F} contains even and odd powers of $\bm{v}$, including the equilibrium force of \cref{eq:eqf}. We may obtain a different form by subtracting the force with the probe traveling in the opposite direction, \ie, with velocity carrying the opposite sign, $\left<\bm{F}\right>^{(-\bm{v})}_{{\bm{x}}}$, which yields,
\begin{equation}
\begin{split}
\left<\bm{F}\right>^{\noneq}_{\bm{x}} - 
\left<\bm{F}\right>^{(-\bm{v})}_{{\bm{x}}} = \hspace{3cm} \\ 
-2\beta\int_{0}^\infty 
\left<\bm{F}(t);\bm{F}(0)\right>^\eq_{{\bm{x}}}\dd{t}\cdot\bm{v} 
+ \order{\bm{v}^3}.
\end{split}
\label{eq:GK_BF}
\end{equation}
\cref{eq:GK_BF}, as the equilibrium term of \cref{eq:eqf} has been removed, yields the force related to energy dissipation, which will be addressed in \cref{sec:Op}. \cref{eq:GK_BF} also misses the  second order response, which is of practical advantage when numerically evaluating the left hand side of \cref{eq:GK_BF} in non-equilibrium simulations. 

We now define the friction (dyadic) tensor $\bm{\gamma}$ as a function of $\bm{x}$. From the right hand side of \cref{eq:GK_BF}, it follows that
\begin{equation}
\bm{\gamma}^\eq({\bm{x}}) := \beta\int_{0}^\infty\left<\bm{F}(t);\bm{F}(0)\right>^\eq_{\bm{x}} \dd{t}  ,
\label{eq:fric_coe_eq}
\end{equation}
which uses only thermal equilibrium quantities. In addition, the left hand side of \cref{eq:GK_BF} allows for a \ahum{direct} definition in terms of non-equilibrium quantities
\begin{equation}
\bm{\gamma}({\bm{x}}) := \frac{1}{2}\left(\left<\bm{F}\right>^{(-\bm{v})}_{\bm{x}} -
\left<\bm{F}\right>^{\noneq}_{{\bm{x}}}\right) \frac{\bm{v}}{v^2}  ,
\label{eq:fric_coe_neq}
\end{equation}
where $v$ is the magnitude of $\bm{v}$, and \cref{eq:fric_coe_neq} is to be understood as the dyadic tensor product of $\bm{F}$ and $\bm{v}$. For sufficiently small $v$, \cref{eq:fric_coe_eq,eq:fric_coe_neq} are expected to coincide. Note that using \cref{eq:fric_coe_neq} requires two independent non-equilibrium simulations to be performed: one with the probe moving with velocity $\bm{v}$, the other with the probe moving with velocity $-\bm{v}$.

\section{Molecular Dynamics}\label{sec:MD}

\begin{figure}
\centering
\includegraphics[scale=1]{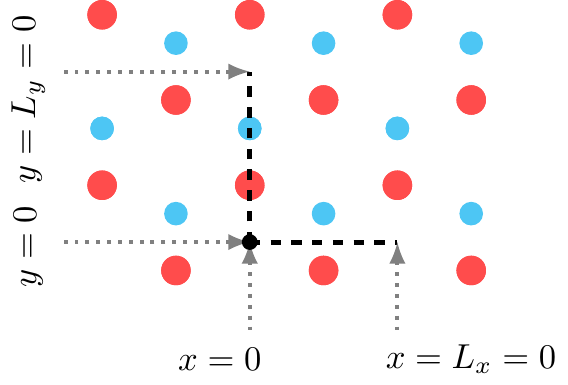}
\caption{Top view on the crystal surface: Red circles represent the topmost layer of the substrate particles, and the smaller cyan circles the layer directly below. Note the hexagonal structure characteristic of the $(111)$-plane. $L_x=\sigma$ and $L_y=\sqrt{3}\sigma$ are the lengths of periodicity in the $x$- and $y$-direction, respectively. The black dot denotes the origin of the coordinate system with regard to which the position of the probe particle is measured, with the corresponding sliding paths considered in this work indicated as dashed lines.}
\label{fig:def_origin}
\end{figure}

The substrate is modeled as an fcc lattice (nearest neighbor distance $\sigma$), prepared by appropriately stacking $(111)$-planes in the vertical $z$-direction. Each single $(111)$-layer corresponds to a hexagonal lattice oriented with respect to the lateral dimensions $x$ and $y$ as shown in \cref{fig:def_origin}. Periodic boundary conditions are applied in the lateral directions. We simulate a total of five vertical layers, each layer containing 144 atoms, the lateral extensions being set as close to square as possible (as allowed by the periodic boundaries). Each lattice site contains a substrate atom (mass $m$) which is connected to its nearest neighbors by permanent bonds. The energy of a single bond features harmonic and anharmonic terms, and is given by,
\begin{equation}\label{eq:bond}
  U^\mathrm{bond} (r) = \sum_{n=2}^4 \alpha_n k (r/\sigma-1)^n  ,
\end{equation}
where $r$ is the distance between the two atoms participating in the bond, stiffness parameter $k$, and constants $\alpha_2=1,~\alpha_3=-7,~\alpha_4=31$ (these constants stem from a Taylor expansion of a regular Lennard-Jones potential around its minimum).

The probe particle is placed on top of the substrate at the position given in the respective graphs (as the probe particle moves with a constant velocity for the simulations to follow, the probe mass is irrelevant). The initial lateral position of the probe is always the high-symmetry position $(x=0, y=0)$ of \cref{fig:def_origin}. The initial height of the probe, as measured from the top substrate layer, is denoted $z$. The probe interacts with the substrate atoms via a short-ranged, purely repulsive, Lennard-Jones pair potential~\cite{10.1002/0471428019.ch5}
\begin{equation}
  U^\mathrm{LJ}( r ) = \begin{cases} 4\epsilon \left[
  \left(\frac{2\sigma}{r}\right)^{12} - \left(\frac{2\sigma}{r}\right)^{6} \right] 
  + \epsilon, & r \le r_c, \\
  0, & \mbox{otherwise},
\end{cases}
\label{eq:WCA}
\end{equation}
with the cutoff $r_c = 2^{7/6} \sigma$ at the distance where the full potential would have its minimum; the additive constant ensures that $U^\mathrm{LJ}(r)$ is always positive. Note that this interaction sets the effective diameter of the probe to $2\sigma$, \ie,~twice the substrate nearest neighbor distance. In the analysis to follow, we also consider the {\it weakly coupled limit} (WCL). In this limit, the probe interacts with the substrate atoms exactly as in \cref{eq:WCA}, but with the implied force contribution {\it excluded} from the substrate atoms. In the WCL, the probe thus acts as a test particle, not affecting the substrate dynamics in any way.

To control the temperature $T$ in our simulations, a Langevin 
thermostat~\cite{10.1103/physrevb.17.1302, 10.1142/s0129183191001037,apostoli17a} is applied to each substrate atom (but not to the probe particle). To the forces acting on these particles, damping and stochastic terms are added,
\begin{equation} 
\label{eq:lan}
 \bm{F}_{\rm lan} = -(m / \tau_{\rm lan}) \, \bm{v}_{\rm sub} 
 + \sqrt{ \frac{24k_\mathrm{B}Tm}{\tau_{\rm lan}\Delta t} } \, \bm{s}.
\end{equation}
Here, $\tau_{\rm lan}$ is the Langevin relaxation time, $m$ the mass of a single substrate atom, $\bm{v}_{\rm sub}$ the instantaneous velocity of the substrate atom, $\Delta t$ the MD integration time step, $k_\mathrm{B}$ the Boltzmann number, and $\bm{s}$ a three-dimensional vector with components uniformly drawn from $[-0.5,0.5]$. As was shown in \olcite{10.1142/s0129183191001037}, uniform and Gaussian random numbers are both suitable for Langevin thermostatting; we use the former due to a (slight) efficiency gain. We emphasize that the use of the Langevin thermostat as described above is by no means the only possibility. One could also have used so-called stochastic boundaries, where the Langevin thermostat is applied exclusively to the boundaries of the system~\cite{10.1088/0953-8984/22/7/074205, 10.1103/physrevb.82.081401, PhysRevB.100.094305}. Our view is that, by changing the thermostat details, one changes the equilibrium dynamical properties of the system, \ie{}, the type of material that is being studied.

In what follows, we use dimensionless (Lennard-Jones) units throughout: $\epsilon = \sigma = m = k_\mathrm{B} \equiv 1$. The bond stiffness in \cref{eq:bond} is set to $k=1000\epsilon$. This large value ensures that the substrate maintains its crystalline structure in the presence of the probe. The dynamics of the substrate particles is obtained via micro-canonical (NVE) integration, using time step $\Delta t = 0.001$, except for the bottom layer of substrate atoms, which are kept frozen (in order to spatially anchor the substrate). All simulations were performed using the software package LAMMPS~\cite{10.1006/jcph.1995.1039}. The Langevin thermostat operates at temperature $T=0.15$, the corresponding relaxation time is set to $\tau_{\rm lan} = 100\Delta t = 0.1$. A finite relaxation time  limits the maximum phonon lifetime to better capture a realistic solid, where these lifetimes are also limited due to defect scattering, electron-phonon coupling ~\cite{10.1021/acs.jpclett.7b03182, 10.1103/physrevlett.116.217601,panizon18a,robbins00a}, and other effects not explicitly included in our simulations. We refer the interested readers to Ref.~\cite{panizon18a} for a detailed analysis of phonon excitation.

To determine the friction tensor elements from equilibrium simulations, \cref{eq:fric_coe_eq} is used. For example, to obtain $\gamma_{xx}$ at the position $\bm{x}=(0,0,1.2\sigma)$ of \cref{fig:def_origin}, the probe is held fixed at that position, and the force component $F_x(t)$ acting on the probe in the $x$-direction is recorded as a function of time; appropriately Fourier transforming the signal $F_x(t)$ yields the force-force covariance (autocorrelation) function, whose time-integrated value equals to, following \cref{eq:fric_coe_eq}, the coefficient $\gamma_{xx}$. By performing similar measurements at regularly spaced probe positions (typically 50 positions per lattice period) along the $x$-direction, the spatially resolved friction coefficient is obtained. In a similar fashion, the coefficient $\gamma_{yy}$ follows from the signal $F_y(t)$, where now the probe samples positions along the $y$-direction.

Upon sliding, we use \cref{eq:fric_coe_neq} to determine the friction tensor. As a concrete example, consider the sliding path through $(0,0,1.2\sigma)$ of \cref{fig:def_origin} running in the positive $x$-direction with velocity $v$. Each time the probe visits the position $\bm{x}$, the corresponding force average $\avg{\bF}^{(\bV)}_{\bX}$ is updated, with positions $\bX$ chosen equally spaced along the lattice period (again 50 positions, with $\avg{\bF}^{(\bV)}_{\bX}$ thus corresponding to a binning average; since we slide over a crystalline surface, there is a well-defined spatial period, and so the average can be collected in one simulation run, by expressing the probe position modulo the spatial period). A second, statistically independent simulation, is performed next, where the sliding proceeds in the negative $x$-direction. Substituting the measurements of both simulations in \cref{eq:fric_coe_neq} yields the friction tensor.

Before taking any measurements, $\num{2e5}$ MD steps are discarded to allow for thermalization. A single measurement then uses $\num{2.5e4}$ MD steps; results presented show an average over 2000 independent measurements (\ie,~different thermostat random numbers). In our simulations, two sliding paths for the probe are considered, both starting in $(0,0,z)$ of \cref{fig:def_origin} at height $z$ above the top substrate layer. One path then proceeds along the $x$-direction, the other along the $y$-direction. Note the different periodicity: $\sigma$ and $\sqrt{3} \sigma$, for $x$ and $y$, respectively.

\section{Results}\label{sec:results}

\begin{figure}
\centering
\includegraphics[scale = 1.]{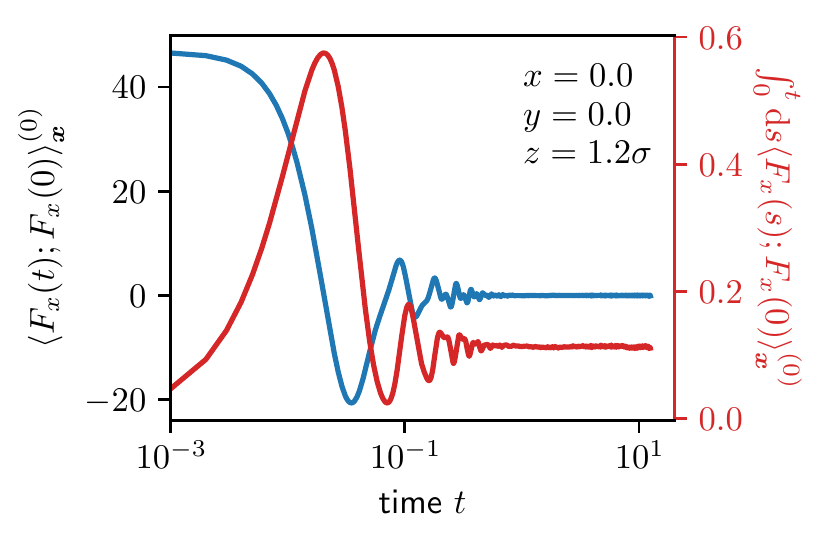}
\vspace{-0.5cm}
\caption{The force-force covariance $\avg{F_x(t);F_x(0)}^\eq_{{\bm{x}}}$ of the probe above the crystal surface (blue curve), and the running time integral of the force-force covariance (red curve).}
\label{fig:ffc_eq_measure}
\end{figure}

\begin{figure}
\subfloat{\centering
\includegraphics[scale = 1]{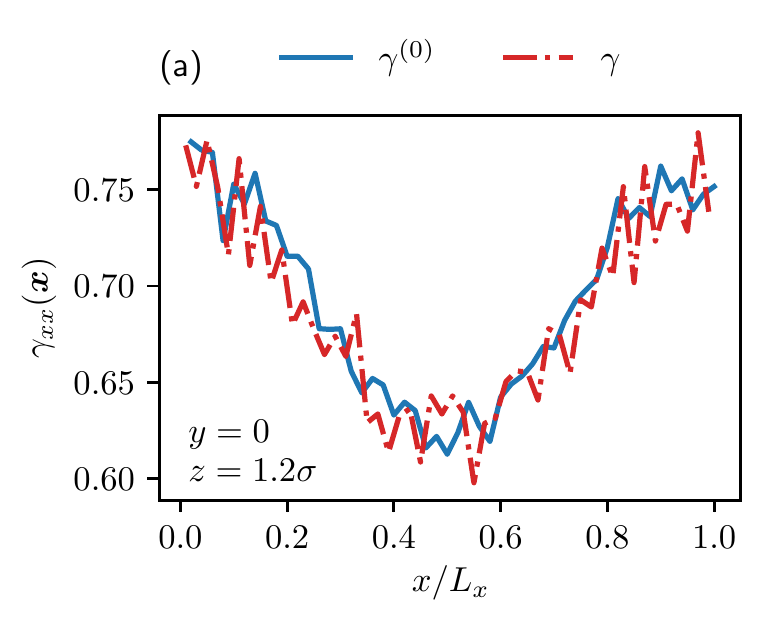}
\label{fig:X_noneq_force}}
\vspace{-0.5cm}
\subfloat{\centering
\includegraphics[scale = 1]{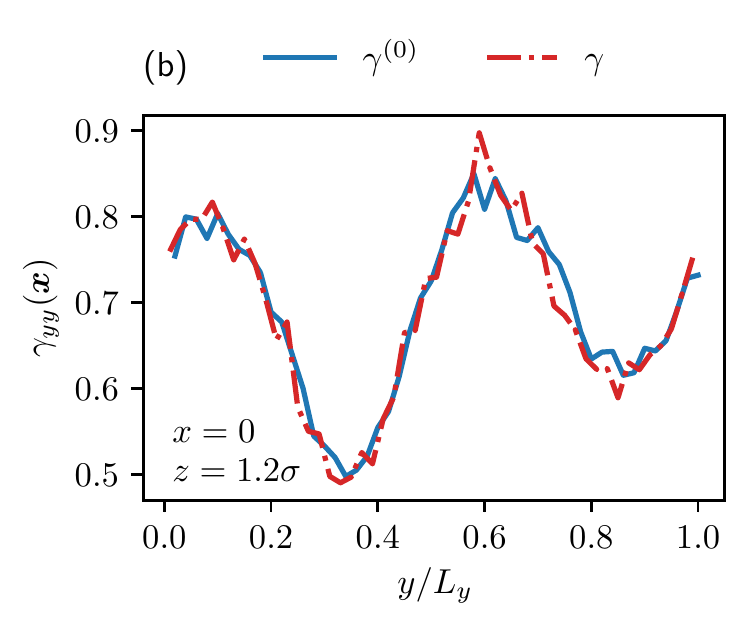}
\label{fig:Y_noneq_force}}
\vspace{-0.5cm}
\caption{The comparison of the equilibrium friction coefficient 
$\gamma^0_{ii}(\bm{x})$ (the blue line) against the non-equilibrium counterpart 
$\gamma_{ii}(\bm{x})$ (the orange dashed line) when the probe moves in (a) $x$- 
and (b) $y$-direction, respectively. Each non-equilibrium friction coefficient 
is averaged over velocities  $v\in[0.1,1.0]$ in units of $\sigma / \Delta t$.}
\label{fig:noneq_force}
\end{figure}

\begin{figure}
\centering
\includegraphics[scale = 0.95]{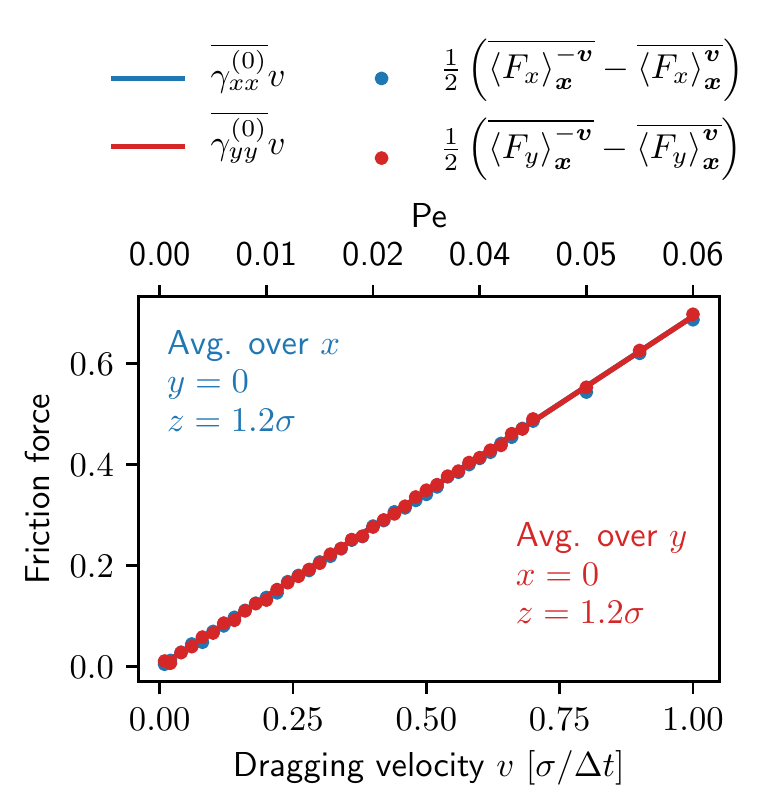}
\vspace{-0.5cm}
\caption{Positionally averaged equilibrium (the solid lines) and non-equilibrium (the dots) friction forces as a function of sliding velocity ($v$) and P\'eclet number (Pe). The blue and the red color codes indicate the $x$- and $y$-direction, respectively. The overline notation denotes the positional average, \eg, $\overline{\gamma^\eq_{ii}} = \frac{1}{L_i}\int_{0}^{L_i} \gamma^\eq_{ii}(\bm{x})\dd{x_i}$.}
\label{fig:fric_vel}
\end{figure}

\subsection{Validation: friction from equilibrium properties}

We first verify if the equilibrium friction measure, \cref{eq:fric_coe_eq}, agrees with \cref{eq:fric_coe_neq} obtained in a non-equilibrium simulation. The essential part of \cref{eq:fric_coe_eq} is the force-force covariance function, which we exemplarily show in~\cref{fig:ffc_eq_measure}, for the $x$-component of forces, with the probe placed at the specific position $\bm{x} = (0,0,1.2\sigma)$. Due to the under-damped dynamics of the substrate particles, this function reveals oscillatory behavior, which decays to zero on a timescale $T^\mathrm{R}$. This timescale allows us to define the P\'eclet number as follows,
\begin{equation}
\mathrm{Pe} := \frac{v T^{\mathrm{R}}}{\sigma},
\label{eq:Pe}
\end{equation}
which compares the relaxation time $T^R$ to the time it takes for the moving probe particle to cover a distance of one lattice constant $\sigma$. If $\mathrm{Pe} \ll 1$, on the time scale of the lattice dynamics $T^R$, the probe moves only a negligible distance, implying that higher order terms in \cref{eq:fric_coe_eq} can safely be neglected. \Cref{fig:ffc_eq_measure} also shows the running value of the time integral of the force-force covariance, which, in the long time limit, yields the component $\gamma_{xx}$ of the friction tensor at the given position (up to a factor of $\beta$). Fully spatially-resolved information about the dissipation behavior, in the linear response regime, may now be obtained by collecting the entries of the friction tensor at different positions $\bm{x}$, as was explained in \cref{sec:MD}.

For the path starting at the above specified point $(0,0,1.2\sigma)$ and proceeding in the $x$-direction, the spatially-resolved friction coefficient $\gamma_{xx}$ is shown in \Cref{fig:noneq_force}(a). The graph shows a pronounced dependence of $\gamma_{xx}$ on $x$, varying between $\sim 0.6$ and $\sim 0.8$. In this graph, we also show the result as obtained using the non-equilibrium \cref{eq:fric_coe_neq}, where, in favor of statistics, we have averaged over sliding velocities in the range $v \in [0.1,1.0]$. 

We note excellent agreement, validating~\cref{eq:fric_coe_eq}, highlight that friction measurements, at low sliding velocities, do not require explicit non-equilibrium simulations to be performed. For completeness, \Cref{fig:noneq_force}b) shows the coefficient $\gamma_{yy}$ for sliding in the $y$-direction, with the sliding path again starting at the point $(0,0,1.2\sigma)$. In this case, an even more pronounced spatial dependence is revealed, with $\gamma_{yy}$ varying between $\sim 0.5$ and $\sim 0.85$, the implications of which are to be discussed in~\cref{sec:Op}.

Note also that, while the curve
in~\cref{fig:noneq_force}a) reveals an inherent symmetry at $x=L_x/2$, no such symmetry for the curve in~\cref{fig:noneq_force}b) is observed. The reason is obvious from the position of the substrate atoms in the layer directly below the top surface layer, see \cref{fig:def_origin}.

\Cref{fig:fric_vel} shows the same data as \cref{fig:noneq_force}, but now spatially averaged over each sliding path, as a function of the sliding velocity $v$. We note once more the excellent agreement between \cref{eq:fric_coe_eq,eq:fric_coe_neq}. While the lower axis gives velocity in simulation units, the upper axis provides the P\'eclet number defined in~\cref{eq:Pe}. For the range shown, the P\'eclet number is indeed small compared to unity, and the friction force is a linear function of velocity. Coincidentally, for the chosen parameters, the spatially-averaged curves for the two directions in \cref{fig:fric_vel} nearly agree. For other parameters (see~\cref{sec:Op} below),  this is not the case.

\subsection{Analyzing the positional dependence}

\begin{figure*}
\centering
\includegraphics[scale = 1]{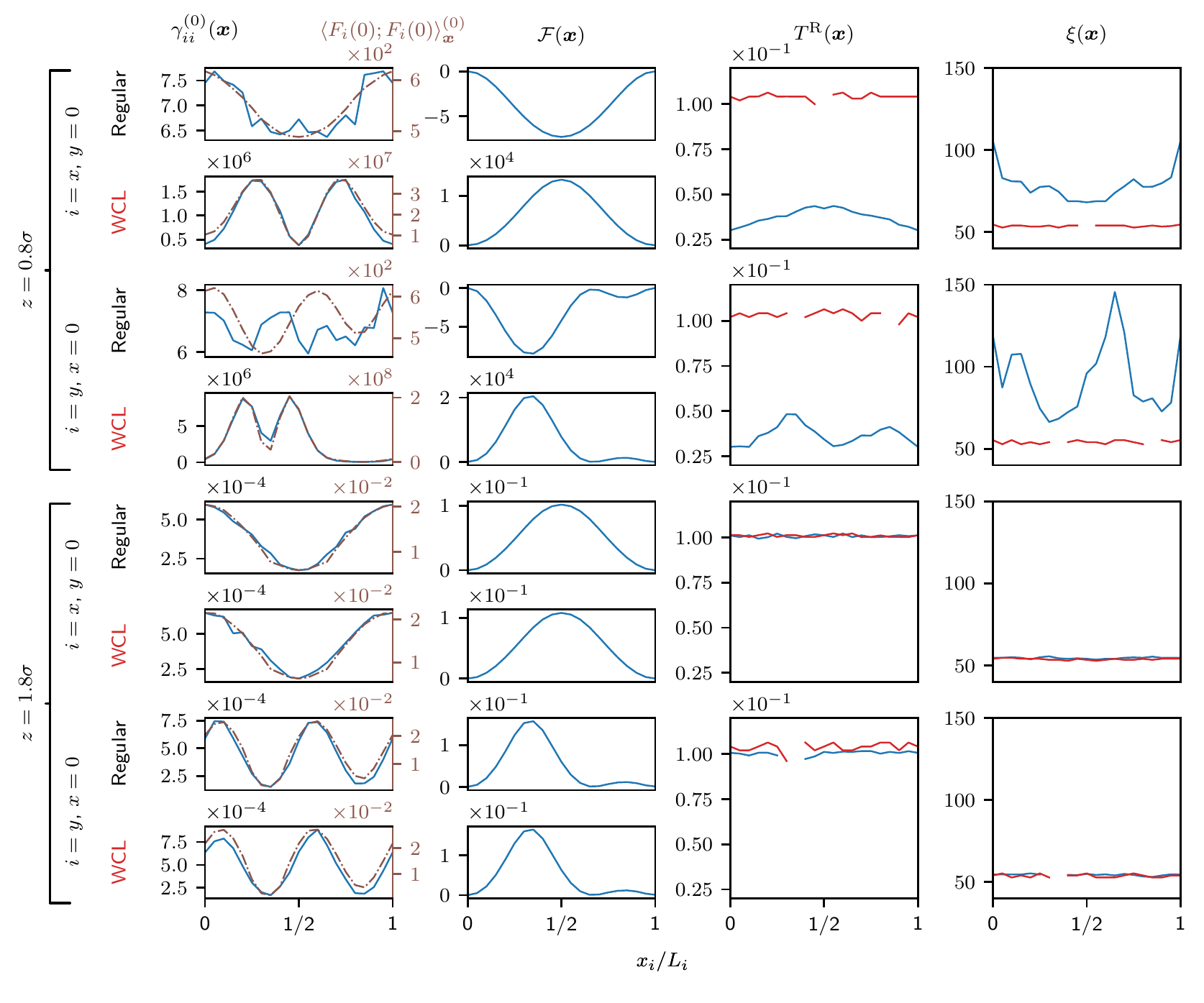}
\vspace{-0.5cm}
\caption{Spatially resolved frictional properties of the regular and weakly-coupled (WCL) probe for two heights $z$ above the substrate upper layer. Column 1: Friction coefficient $\gamma_{ii}^\eq (\bm{x})$, and force-force covariance $\avg{F_i(0); F_i(0)}^\eq_{\bm{x}}$ measured at equal times. Column 2: Free energy $\mathcal{F}(\bm{x})$. Column 3: Relaxation time $T^{\mathrm{R}}(\bm{x})$. Column 4: Oscillation frequency $\xi(\bm{x})$. For columns 3 and 4, blue (red) refers to the regular (WCL) probe. For some positions $\bm{x}$, the force-force covariance does not follow~\cref{eq:covariance_analytic}, in which case $T^{\mathrm{R}}(\bm{x})$ and $\xi(\bm{x})$ remain undetermined.}
\label{fig:spatial_dep}
\end{figure*}

How can the spatial dependence of friction be understood? Understanding the dependence requires analyzing the time dependent force-force covariance of \cref{eq:fric_coe_eq}, $\avg{F_i(t);F_i(0)}^\eq_{\bm{x}}$, shown in \cref{fig:ffc_eq_measure} for the component $i=x$. Since the shape of the curve in \cref{fig:ffc_eq_measure} is reminiscent of a damped harmonic oscillator, we attempt to fit it via the solution $f$ of such an oscillator,
\begin{equation}
f(t) = f_0(\bm{x})\cos[\xi(\bm{x}) t]e^{-\frac{t}{T^\mathrm{R}(\bm{x})}}.
\label{eq:covariance_analytic}
\end{equation}
Here, we introduced a frequency $\xi$; the relaxation time $T^\mathrm{R}$ is defined as above. Note that all parameters in \cref{eq:covariance_analytic} are allowed to depend on the probe position $\bm{x}$. Assuming that \cref{eq:covariance_analytic} is a valid description of the force covariance, we obtain, by integrating, the following expression for the friction coefficient,
\begin{equation}
\gamma^{\eq}_{ii}(\bm{x}) \simeq \frac{\beta T^\mathrm{R}(\bm{x})}{
1+(T^\mathrm{R}(\bm{x})\xi(\bm{x}))^2}\avg{F_i(0);F_i(0)}^\eq_{\bm{x}}
\label{eq:fric_analytic}
\end{equation}
with $i\in\{x,y\}$. The denominator of \cref{eq:fric_analytic} is, for the chosen parameters, dominated by the second term, as $T^\mathrm{R}(\bm{x})\xi(\bm{x}) \approx 5$, which is another indication of the system's underdamped nature.

How do the different terms in \cref{eq:fric_analytic} depend on $\bm{x}$? To address this question, we first consider the {\it weak coupling limit} (WCL), which is formally obtained by letting the potential prefactor $\epsilon$ in \cref{eq:WCA} go to zero, as detailed in \cref{sec:MD}. The friction coefficient is then of order $\epsilon^2$, while the motion of the substrate particles then becomes independent of the presence of the probe. In order to be able to compare the numbers for different cases, we achieve the weak coupling limit by keeping $\epsilon$ finite in \cref{eq:WCA} and removing {\it by hand} the force arising from \cref{eq:WCA} from the equation of motion of the substrate particles.

\Cref{fig:spatial_dep} shows the resulting dependence of the parameters in \cref{eq:fric_analytic} on $x$ and $y$ (regarding the components $xx$ and $yy$ of the friction tensor $\gamma$), for two different values of the height $z$. We note the intriguing observation that, in the weakly-coupled limit, neither $\xi$ nor $T^\mathrm{R}$ reveal any pronounced spatial dependence. Moreover, their values are the same for both the $xx$ and $yy$ components of the friction tensor. The relaxation time $T^\mathrm{R}$ coincides with the Langevin relaxation time $\tau_\mathrm{lan}$ of the MD thermostat of \cref{eq:lan}, reflecting the equilibrium properties of the crystal which are unaffected by the slider-crystal interaction. The consequence is that the dependence of $\gamma$ on position and direction mostly stems from the static (equal-time) force-force covariance $\avg{F_i(0);F_i(0)}^\eq_{\bm{x}}$. This is demonstrated in the first column of the figure, where $\gamma$ and $\avg{F_x(0);F_x(0)}^\eq_{\bm{x}}$ are both shown in one graph, with very good agreement (naturally, up to an overall prefactor).

The other set of rows in \cref{fig:spatial_dep} shows the {\it regular} case, where the substrate particles feel the presence of the probe. For large values of $z$, the regular and weakly-coupled cases are very similar, as expected; if the distance between probe and and substrate is large, the substrate hardly notices the probe. For the regular probe at smaller distances, however, $\xi$ and $T^\mathrm{R}$ do depend on the position of the probe. This is understood as the presence of the probe now \textit{deforms} the crystal, making it locally anisotropic and inhomogeneous~\cite{10.1002/0471428019.ch5}. Probably coincidentally, when sliding in the $x$-direction, $\gamma$ and $\avg{F_x(0); F_x(0)}^\eq_{{x}}$ still agree quite well, because the $x$ dependence of $\xi$ and $T^\mathrm{R}$ seem to almost cancel in \cref{eq:fric_analytic}. It is also worth noting that there is no apparent relation between the friction coefficients and the free energy landscape, as shown in the second column of \cref{fig:spatial_dep}.

Note that some positions $\bm{x}$ exist where the force-force covariance does not follow~\cref{eq:covariance_analytic}, resulting in undetermined relaxation time $T^\mathrm{R}$ and oscillation frequency $\xi$.

\subsection{Energy loss}\label{sec:Op}

\subsubsection{General}

Are there practical consequences of the position and space dependence of $\gamma$? In this section, we analyze how the energy dissipation of the moving probe can be minimized under the constraint of traveling
between two points in given time $\tau$. This analysis uses the friction tensor found from equilibrium fluctuations. To reduce the number of possibilities, we require the
height $z$ to be fixed and not varied. During such motion, the probe is subject to the force given in
\cref{eq:GK_F}. The first term is the gradient of free energy, so that this force, under motion of the
probe, causes a \emph{reversible} change of free energy (\eg, by reverting the
direction of probe motion). The second term in \cref{eq:GK_F}, the friction force, is related to dissipated
energy (it can not be recovered). To leading order in $v$, the dissipated energy $Q$ is thus given by
\begin{equation}
\begin{split}
Q &= \int   \bm{v}(\bm{x})\cdot \bm{\gamma}^{(0)}(\bm{x})\cdot\dd{\bm{l}}\\
&= \int_{0}^{\tau} \bm{v}(t)\cdot
\bm{\gamma}^0(\bm{x}(t))\cdot \bm{v}(t)\dd{t},
\label{eq:Eloss}
\end{split}
\end{equation}
with the path tangential increment $ d\bm{l}$. In \cref{eq:Eloss}, we allowed $\bm{v}$ to be a function
of space, so that it is also a function of time. While this is against our initial assumption entering
\cref{eq:GK_F},  \cref{eq:GK_F} remains valid as long as the velocity of the probe changes little during
the relaxation time $T^R$. This gives rise to another dimensionless number $T^R \partial_t \log v_i$,
which we assume small in the following for any velocity component $i$.

Equation \eqref{eq:Eloss} is reminiscent of the classical action of a free particle with $\bm{\gamma}^{(0)}$ playing the role of a space dependent tensorial mass (apart from units).  
Applying the Euler-Lagrange equation leads to the following equations of motion for extremized dissipation $Q$ (suppressing the arguments for the sake of brevity)
\begin{equation}
\bm{\gamma}^\eq \cdot \bm{\dot{v}} = \left[(\bm{v}\cdot\nabla)\bm{\gamma}^\eq - \frac{1}{2}\bm{v}\cdot(\nabla\bm{\gamma}^\eq)\right]\cdot\bm{v}.
\label{eq:eqofmotion}
\end{equation}

\subsubsection{Comparing paths}

\begin{figure}
\centering
\includegraphics[scale=1]{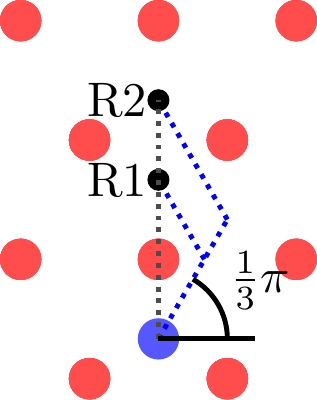}
\caption{Different trajectories meeting at the rendezvous R1 and R2. Due to the symmetries  of the crystal, the detouring trajectories (blue) are equivalent to moving into the $x$-direction.}
\label{fig:bending}
\end{figure}

We start with keeping the magnitude of velocity $v$ time independent, and compare the  exemplary paths shown in~\cref{fig:bending}, with the rendezvous points R1 and R2 reached at time $\tau$. Because the fcc (111)-plane is invariant under rotation by $\frac{1}{3}\pi$, the diagonal paths are physically equal to paths along the $x$-direction as \eg, shown in \cref{fig:noneq_force}a). This allows us to label the lower axis of \cref{fig:Eexp} with $x$.

The solid lines in~\cref{fig:Eexp} represent the  energy loss per $k_\mathrm{B}T$ as a function of
trajectory length along the corresponding path. Because the probe taking either path is to arrive at each rendezvous point at the
same time, the probe taking the detour naturally travels faster. In terms of energy loss, it
thus has the double disadvantage of a longer path and a larger speed. Nevertheless, in the given case, the longer
trajectory is beneficial when meeting at the rendezvous point
R1. When meeting at R2, the two trajectories dissipate about the same amount of energy. This exemplifies the nontrivial dependence of dissipated energy for different paths.

\subsubsection{Optimizing magnitude of velocity}
\begin{figure}
\centering
\includegraphics[scale = 1.]{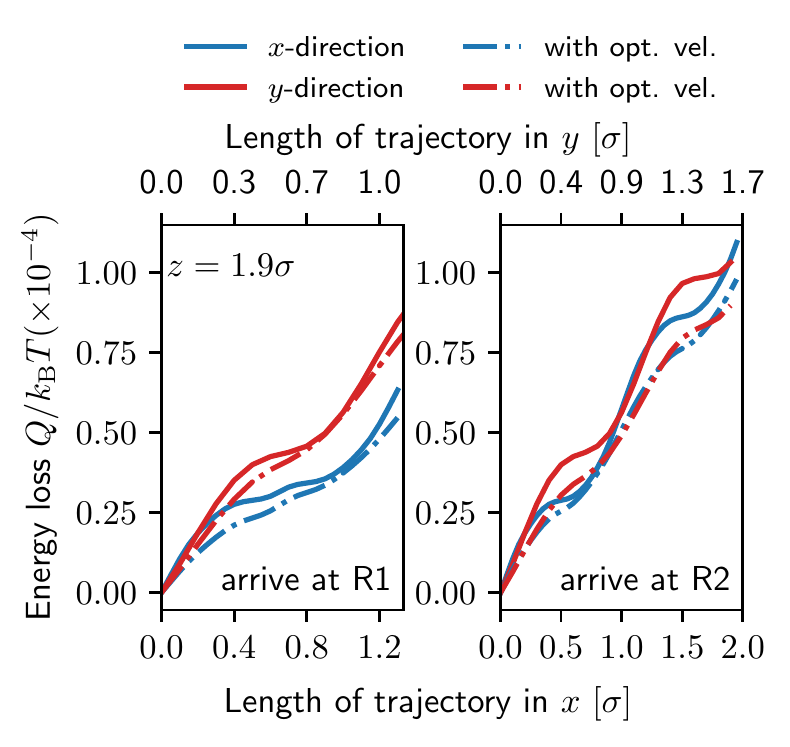}
\vspace{-0.5cm}
\caption{The energy loss with a constant velocity (the solid lines) and with an optimized velocity
protocol (the dashed lines) in the $x$- and $y$-direction (the blue and the red, respectively) as a
function of trajectory length when meeting at R1 (right) and R2 (left), see \cref{fig:bending}. The average velocities in the
$x$- and $y$-direction are 0.12, and  0.1, respectively.}
\label{fig:Eexp}
\end{figure}
\begin{figure}
\subfloat{\centering
\includegraphics[scale =.95]{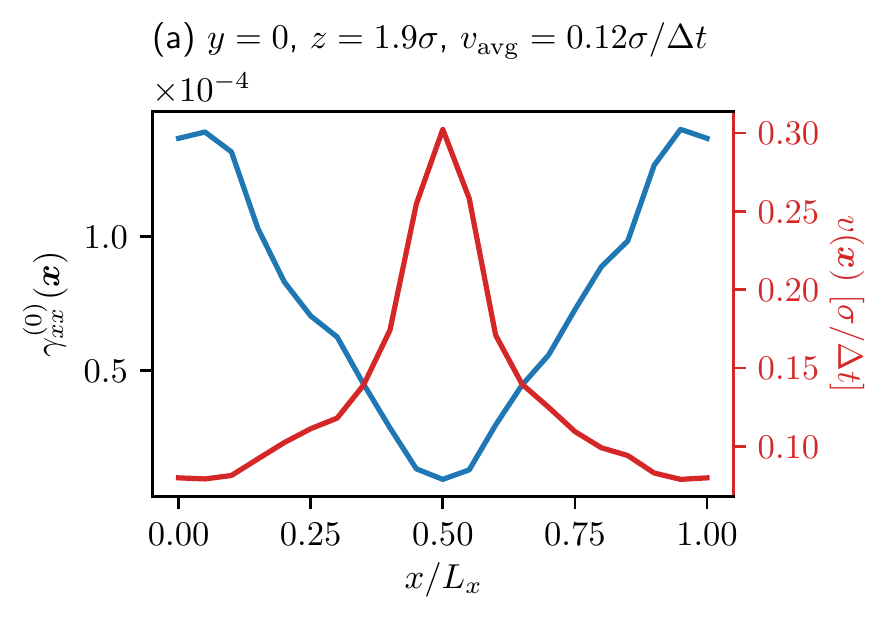}}
\vspace{-0.5cm}
\subfloat{\centering
\includegraphics[scale =.95]{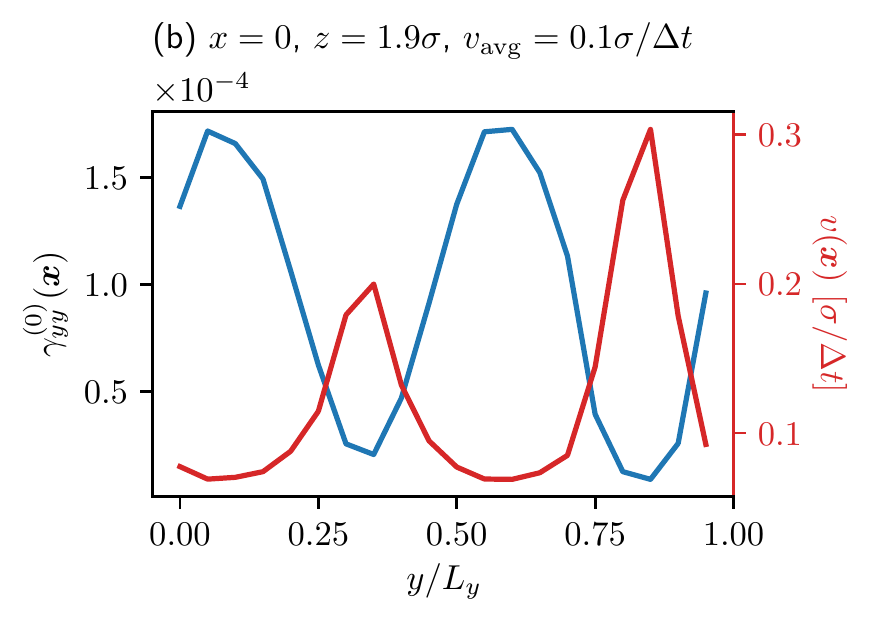}}
\vspace{-0.5cm}
\caption{The friction coefficients $\gamma^{(0)}_{ii}(\bm{x})$ and the corresponding energy loss
optimizing velocity protocols $v(\bm{x})$ when the probe moves in (a) $x$- and 
(b) $y$-direction.}
\label{fig:vel_pro}
\end{figure}
\begin{figure}
\centering
\includegraphics[scale = 1.0]{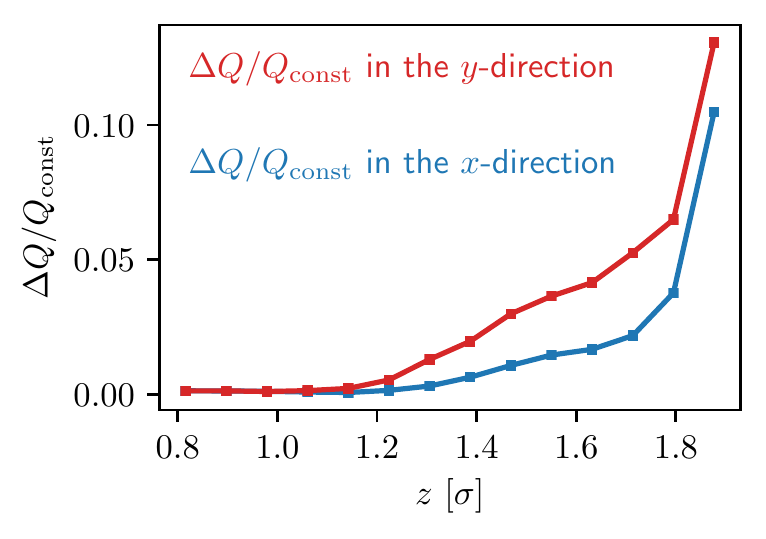}
\vspace{-0.5cm}
\caption{The saving in energy as a function of height when the probe moves in 
$x$- and $y$-direction provided that the probe arrives at R2 
in~\cref{fig:bending}.}
\label{fig:Esaving}
\end{figure}
Can the energy loss be optimized for a given path by varying the {\it magnitude} of velocity? To answer this, we solve~\cref{eq:eqofmotion} for a fixed direction of velocity subject to the condition that the probe moves from $x(0)$ to $x(\tau)$ in time $\tau$ while keeping $y$ and $z$ fixed (without loss of generality we present the following equations with the $x$-entry),
\begin{equation}
v_x(\bm{x}) = \frac{C_x(\tau)}{\sqrt{\gamma^{(0)}_{xx}(\bm{x})}}
\end{equation}
with
\begin{equation}
C_x(\tau) = \frac{1}{\tau}\int_{x(0)}^{x(\tau)}\sqrt{\gamma^{(0)}_{xx}(\bm{x})}\dd{x}.
\end{equation}
The optimizing velocity protocol exploits the positional dependence of the friction coefficient by moving slowly  when the friction coefficient is high and \textit{vice versa} as seen in~\cref{fig:vel_pro}.  The optimized energy loss is presented as the dashed lines in~\cref{fig:Eexp}.

The saving in energy is hence subject to how strongly the friction varies as a 
function of position. More precisely, the maximal amount of possible energy 
saving (compared to the case of constant velocity) is a functional of 
$\gamma_{ii}(x)$. This reads, for instance in the $x$-direction,
\begin{equation}
\frac{\Delta Q}{Q_\mathrm{const}} = 1-
\frac{\left[\int_{x(0)}^{x(\tau)}\sqrt{\gamma^0_{xx}(\bm{x})}\dd{x}\right]^2}
{(x(\tau)-x(0))\int_{x(0)}^{x(\tau)}\gamma^0_{xx}(\bm{x})\dd{x}},
\label{eq:eratio}
\end{equation}
where $\Delta Q:=Q_\mathrm{const}-Q_\mathrm{opt}$ is the difference of the energy loss with a constant velocity protocol $Q_\mathrm{const}$ and the optimized energy loss $Q_\mathrm{opt}$. \Cref{eq:eratio} displays that a high variance of $\sqrt{\gamma^0_{xx}(\bm{x})}$ in
$\bm{x}$ results in large savings from optimization. As such, 
\cref{fig:Esaving} shows that one can save more energy when the probe moves in 
the $y$-direction within the given range of height. This is because the 
$yy$-entry of the friction tensor is a more pronounced function of $\bm{x}$ 
along that path.

We finally note, as a curiosity, that the velocity protocol of
$
v_x(\bm{x}) = \frac{C'_{x}(\tau)}{\gamma^0_{xx}(\bm{x})}
$
with
$
C'_{x}(\tau) = \frac{1}{\tau}\int_{x(0)}^{x(\tau)} \gamma^0_{xx} \dd{x},
$
yields the exact same energy loss as a protocol with  a constant velocity.  A short proof is given in~\cref{appn:inverse}.

\section{Conclusion}\label{sec:conc}

We have employed Green-Kubo relations to study sliding friction of a probe particle moving over the surface of a model fcc crystal in the limit of low sliding speed, using molecular dynamics simulations. The Green-Kubo relations naturally facilitate obtaining the friction spatially and directionally resolved at atomic scales. For the model simulated by us, we find a pronounced spatial and directional dependence of the friction tensor elements. We have rationalized these findings by considering the limit of weak coupling, where the probe hardly disturbs the motion of substrate particles. In this limit, the local variation of friction is well captured by the {\it equilibrium} force covariance at equal times. This observation, whose origin lies in the homogeneity of oscillation frequency and relaxation time, is worth noting, as it relates a {\it dynamic} observable to {\it static} properties of the system. This analysis suggests that, in the weakly coupled limit, the spatial and directional dependence of sliding friction force can be obtained using a thermal equilibrium Monte Carlo simulation, since here the details of the dynamics appear irrelevant. Giving up the limit of weak coupling, where the probe notably disturbs the surface particles, we observe that also the oscillation frequency and relaxation time of fluctuations depend on the position of the probe, so that the spatial dependence of friction becomes more complicated and depends on dynamical details. Intriguingly, the most easily accessible static property, the (free) energy landscape, is {\it not} a good indicator for the spatial dependence of the friction coefficient. 

The spatial and directional dependence of friction can be used to minimize energy loss, \eg, under the constraint of traveling between two points in a certain amount of time. This amounts to an Euler Lagrange equation for the particle in terms of the space dependent friction tensor. Specifically, in some cases, taking a \ahum{detour} can be of advantage! It also concerns the speed, so that it is beneficial to move fast in regions of low friction and \textit{vice versa}. Quantitatively, the optimal sliding
protocol implements a speed that is proportional to the inverse square root of the friction
coefficient. 

Future work will extend the given model to include extra degrees of freedom whereby the probe is being dragged by a spring, reminiscent of an AFM setup~\cite{gauthier99a}. In addition, we aim to study cases of macroscopic surface inhomogeneity, such as for amorphous solids, or for surfaces with more realistic energy landscapes. Notably, the presented approach is valid for such cases as well, as long as the sliding velocity remains sufficiently slow. Extending our findings to the case of larger probe velocities, \ie, beyond the nonlinear regime, using methods of Ref.~\cite{basu15a}, is also of great interest.

\acknowledgments

We acknowledge support by the German research foundation through the SFB-1073 (RV, Project A01). We thank Cynthia Volkert and Peter Sollich for stimulating discussions.

\appendix

\section{The inverse-gamma velocity protocol}\label{appn:inverse}

Suppose that the velocity protocol is given as
\begin{equation}
\dv{x}{t} = \frac{C'_{x}}{\gamma^0_{xx}(\bm{x})}
\end{equation}
with an arbitrary constant $C'_{x}$.
To determine the constant $C'_{x}$, we re-arrange the above expression and take integrals on both
sides
\begin{equation}
\begin{split}
\int_{x(0)}^{x(\tau)}\gamma^0_{xx}(\bm{x})\dd{x}& = C'_{x}\int_{0}^{\tau}\dd{t}\\& = C'_{x} \tau.
\end{split}
\end{equation}
This leads us to
\begin{equation}
C'_{x}(\tau) = \frac{1}{\tau}\int_{x(0)}^{x(\tau)} \gamma^0_{xx}(\bm{x}) \dd{x}.
\end{equation}
That is, the constant $C'_{x}$ is uniquely determined by the constraints.
The complete expression for the velocity protocol is thus
\begin{equation}
v_x(\bm{x}) = \frac{1}{\tau\gamma^0_{xx}(\bm{x})}\int_{x(0)}^{x(\tau)} \gamma^0_{xx} (\bm{x'})\dd{x'}.
\end{equation}
Plugging this velocity protocol to the energy loss formula~\cref{eq:Eloss} yields
\begin{equation}
\begin{split}
Q &= \frac{1}{\tau}\int_{x(0)}^{x(\tau)}
\frac{\gamma^{(0)}_{xx}(\bm{x'})}{\gamma^0_{xx}(\bm{x'})}\int_{x(0)}^{x(\tau)}
\gamma^0_{xx} (\bm{x''})\dd{x'}\dd{x''}\\
&= \frac{x(\tau)-x(0)}{\tau}\int_{x(0)}^{x(\tau)}\gamma^0_{xx} (\bm{x''})\dd{x''}\\
&=v_\mathrm{avg} \int_{x(0)}^{x(\tau)}\gamma^0_{xx} (\bm{x''})\dd{x''}.
\end{split}
\end{equation}
It means assigning a velocity protocol that is inversely proportional to the friction coefficient is essentially the
same as assigning a constant velocity protocol.

\bibliography{bibtex}
\bibliographystyle{apsrev4-1}
\end{document}